# Unified Growth Theory Contradicted by the GDP/cap Data

Ron W Nielsen[1]

Environmental Futures Research Institute, Gold Coast Campus, Griffith University, Qld, 4222, Australia

December, 2015

Mathematical properties of the historical GDP/cap distributions are discussed and explained. These distributions are frequently incorrectly interpreted and the Unified Growth Theory is an outstanding example of such common misconceptions. It is shown here that the fundamental postulates of this theory are contradicted by the data used in its formulation. The postulated three regimes of growth did not exist and there was no takeoff at any time. It is demonstrated that features interpreted as three regimes of growth represent just mathematical properties of a single, monotonically-increasing distribution, indicating that a single mechanism should be used to explain the historical economic growth. It is shown that using different socio-economic conditions for different perceived parts of the historical GDP/cap data is irrelevant and scientifically unjustified. The GDP/cap growth was indeed increasing slowly over a long time and fast over a short time but these features represent a single, uniform and uninterrupted growth process, which should be interpreted as whole using a single mechanism of growth.

**Introduction**

Hyperbolic distributions appear to be creating significant problem with their interpretation. They are routinely seen as being made of two distinctly different components, slow and fast, jointed by a transition stage. However these distributions are easy to understand if they are represented by their reciprocal values (Nielsen, 2014) because in this representation the confusing features disappear and hyperbolic distributions are represented by straight lines.

Significantly more difficult problem is to understand the distributions representing the historical Gross Domestic Product *per capita* (GDP/cap) because features, which were already difficult to understand for hyperbolic distributions, are even more confusing. In addition, these distributions cannot be simplified by their reciprocal values and to understand them we have to use a different approach.

Incorrect interpretations of the historical GDP/cap data is a serious problem and the prominent example is the Unified Growth Theory (Galor, 2005a; 2011). Using the reciprocal values of the GDP data, it has been already demonstrated (Nielsen, 2014) that the fundamental postulates of this theory are contradicted by empirical evidence. We shall now demonstrate that the same conclusion can be reached by analysing the GDP/cap data coming from precisely the same source as used in developing this theory.

---

[1]AKA Jan Nurzynski, r.nielsen@griffith.edu.au; ronwnielsen@gmail.com;
http://home.iprimus.com.au/nielsens/ronnielsen.html

The Unified Growth Theory is aimed at explaining the apparent different stages of growth but we shall demonstrate that this explanation is irrelevant because the GDP/cap data follow a *single* trajectory, indicating that they should be explained using a single mechanism. We shall demonstrate that the three regimes of growth did not exist and that there was no takeoff in the economic growth at any time.

**Crude representation of data**

The GDP/cap distributions are frequently displayed in a grossly simplified way by selecting just four strategically-located points (Ashref, 2009; Galor, 2005a, 2005b, 2007, 2008a, 2008b, 2008c, 2010, 2011, 2012a, 2012b, 2012c; Galor and Moav, 2002; Snowdon & Galor, 2008) as shown in the top panel of Figure 1. In this figure we show an example for the world economic growth but similar plots are also used for the regional data. Such displays are strongly suggestive and they serve as a perfect prescription for drawing incorrect conclusions. However, plotting more data in a certain way can be also strongly suggestive and confusing (see the lower panel of Figure 1).

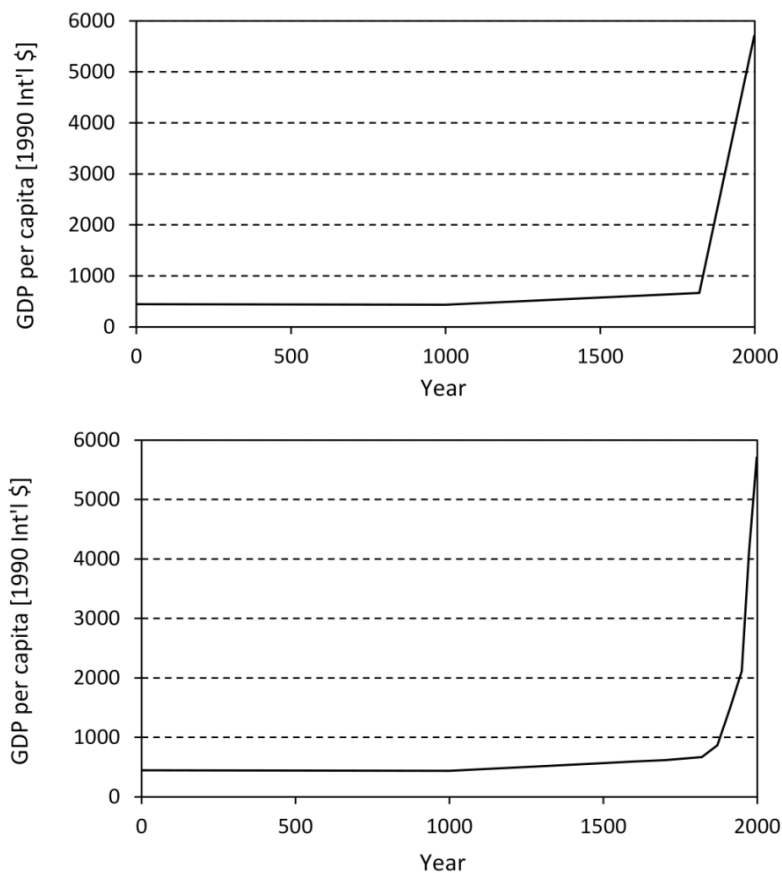

Figure 1. Gross Domestic Product (GDP) per capita (Maddison, 2001) as frequently plotted (Ashref, 2009; Galor, 2005a, 2005b, 2007, 2008a, 2008b, 2008c, 2010, 2011, 2012a, 2012b, 2012c; Galor and Moav, 2002; Snowdon & Galor, 2008) to explain the mechanism of the historical economic growth.

The GDP/cap distributions are already sufficiently confusing even if all data are plotted without joining them by straight lines. They suggest a prolonged epoch of stagnation represented by the low values of the GDP/cap ratio, followed by a rapid takeoff representing an alleged new regime of economic growth governed by a distinctly different mechanism. Such distributions have to be analysed with extra care but plots such as shown in Figure 1 are not helpful because they reinforce incorrect impressions and interpretations.

Impressions can be misleading and every effort should be taken to avoid being guided by their deception. Science is not based on impressions but on rigorous analysis of data. Unified Growth Theory (Galor, 2005a; 2011) describes ideas based on impressions created by such displays as shown in Figure 1 or by quoting certain data without making any effort to analyse them scientifically. Incorrect concepts remain incorrect even if translated into mathematical formulae. It has already been shown that the fundamental concepts of the Unified Growth Theory are contradicted by the historical GDP data (Nielsen, 2014). We shall now show that they are also contradicted by the GDP/cap data.

**Explaining the GDP/cap ratio**

The GDP/cap ratio combines two time-dependent distributions: (1) the time-dependent GDP growth and (2) the time-dependent population growth. In order to understand the GDP/cap distributions we have to understand their two components: the growth of the GDP and the growth of the population.

Over 50 years ago, von Foerster, Mora and Amiot (1960) demonstrated that the world population was increasing hyperbolically during the AD era. The natural tendency of the historical GDP growth (not only global but also regional) is also to follow hyperbolic distributions (Nielsen, 2014, 2015a, 2015b).

Even though a hyperbolic distribution appears to be made of two different components, slow and fast, joined by a transition component, it has been shown (Nielsen, 2014) that such interpretation is based on strongly misleading impressions. Reciprocal values of a hyperbolic distribution describing growth follow a decreasing straight line and it is then obvious that it makes no sense to divide a straight line into arbitrarily selected sections and claim different mechanisms of growth for each section. It also makes no sense to look for a point marking a takeoff on such a monotonically decreasing straight line because a monotonically decreasing straight line remains a monotonically decreasing straight line and there is no justification in selecting a certain point on such a line and claim that there is a change of direction at this point because there is no change of direction.

The first and essential step in trying to understand the GPD/cap data, regional or local is to check whether their two components (GDP and population) follow hyperbolic distributions. Based on the available evidence (Nielsen, 2014, 2015a, 2015b; von Foerster, Mora & Amiot, 1960), they are likely to demonstrate such preference. Consequently, in order to understand the historical GDP/cap data we have to understand the mathematical process of dividing two hyperbolic distributions.

We are going to demonstrate now that the characteristic features of the GDP/cap distributions, which were used in the formulation of the Unified Growth Theory (Galor, 2005a, 2011), represent purely *mathematical properties* of dividing two hyperbolic distributions. They do not represent different socio-economic conditions describing different mechanisms of growth for different perceived sections of these distributions as claimed in the Unified Growth Theory.

Hyperbolic distribution describing *growth* is represented by a *reciprocal* of a linear function:

$$f(t) = (a - kt)^{-1}, \qquad (1)$$

where $f(t)$ is the size of the growing entity, $t$ is the time, and $a$ and $k$ are *positive* constants.

A reciprocal of hyperbolic distribution, $[f(t)]^{-1}$, is represented by a *decreasing* straight line:

$$[f(t)]^{-1} \equiv \frac{1}{f(t)} = a - kt. \qquad (2)$$

Hyperbolic *distributions* should not be confused with hyperbolic *functions* ($\sinh(t)$, $\cosh(t)$, etc). Furthermore, *reciprocal* distribution or functions, $[f(t)]^{-1}$, should not be confused with *inverse* functions, $f^{-1}(t)$. Mathematical symbol for the inverse function, $f^{-1}(t)$, is similar to the mathematical symbol for the reciprocal function, $[f(t)]^{-1}$, but the concepts are different.

In the inverse functions, the roles of variables are inversed. In the reciprocal functions they remain the same. Thus, for instance, for the distribution given by the equation (1), the objective of using its inverse function would be to calculate how the *time* depends on the size of the growing entity. The inverse of the eqn (1) is

$$f^{-1}(t) = \frac{a}{k} - \frac{1}{kt}, \qquad (3)$$

where $t$ is now the size of the growing entity and $f^{-1}(t)$ is the time. For the reciprocal function given by the eqn (2), $t$ is still the time as in the eqn (1). From the eqn (3) we can see that when the size of the growing entity, $t$, increases to infinity, the time, $f^{-1}(t)$, reaches its terminal value of $a/k$.

The characteristic feature of hyperbolic distributions is that they increase slowly over a long time and fast over a short time, escaping to infinity at a certain fixed time $t_s = a/k$, i.e. when the denominator in the eqn (1) approaches its zero value. However, as we have already pointed out and as discussed earlier (Nielsen, 2014), it is a mistake to interpret such distributions as being made of two distinctly different components joined by a transition component. It is one and continuous distribution, which has to be interpreted as a whole. If such a distribution represents a certain mechanism of growth, it is the same mechanism for the whole distribution.

Let us now take two hyperbolic distributions, $f(t)$ and $g(t)$, and let us divide them. Results are presented in Figure 2.

Parameters describing hyperbolic distributions displayed in Figure 2 are: $a = 4.5$ and $k = 2.2 \times 10^{-3}$ for $f(t)$ and $a = 7$ and $k = 3.35 \times 10^{-3}$ for $g(t)$. *These distributions are purely mathematical entities. They have nothing to do with the growth of the population or with the economic growth.* However, they satisfy a simple condition: the singularity of the $f(t)$ distribution occurs earlier than the singularity of the $g(t)$ distribution. For the curves displayed in Figure 2 singularities are at $t_s \approx 2045$ for $f(t)$ and $t_s \approx 2090$ for $g(t)$. The point of singularity for the $f(t)/g(t)$ ratio is, of course, at $t_s \approx 2045$.

When the distribution $f(t)$ is divided by $g(t)$ they produce a distribution, which resembles closely a typical GDP/cap distribution (see the lower panel of Figure 1). The characteristic features of this distribution are a long stage of nearly constant values of the $f(t)/g(t)$ ratio followed by a nearly vertical increase.

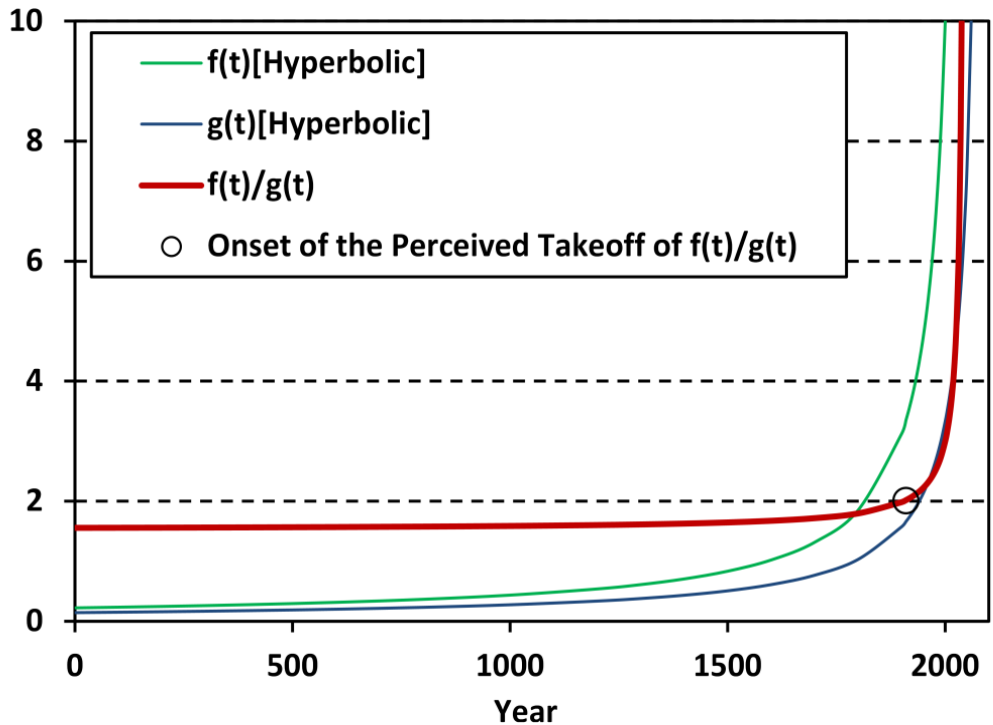

Figure 2. Two, mathematically-defined, hyperbolic distributions, $f(t)$ and $g(t)$, and their ratio $f(t)/g(t)$. The time of the perceived, but non-existent, takeoff is indicated.

It is important to understand that these features are more pronounced than for the corresponding hyperbolic distributions. The nearly horizontal part is flatter and the nearly vertical part is even more vertical. That is why, if the hyperbolic distributions are already so confusing, the distributions representing the ratio of two hyperbolic distributions are even more confusing and their interpretation is even more difficult. They have to be analysed with extra care and their analysis cannot be simplified by using their reciprocal values because the reciprocal of the ratio of two hyperbolic distributions is also a ratio of two hyperbolic distributions. Their analysis is significantly more difficult than the analysis of hyperbolic distributions. They represent a well-concealed trap suggesting strongly the existence of two or even three different components and even the most experienced researcher, who is not familiar with hyperbolic distributions or who is reluctant to accept them because of their singularity, can be easily misguided.

So we can see now that by dividing two, mathematically defined distributions, which have nothing to do with the economic growth, we have generated the fundamental features, which inspired the creation of the Unified Growth Theory (Galor, 2005a, 2011): "the Malthusian Regime" represented by the flat "part," "Sustained-Growth Regime" represented by the steep "part," and the middle "part" which we could call "the Post-Malthusian Regime," containing also a perceived "takeoff," i.e the apparent fast transition from the flat to the steep growth.

These features were generated using purely mathematical, monotonically-increasing distributions and a mathematical division of these distributions. They reflect purely mathematical properties of a *single* distribution representing the $f(t)/g(t)$ ratio. They do not describe different stages of growth. Furthermore, it is clear that these features cannot be attributed uniquely to the GDP/cap distributions. The division of two hyperbolic distributions may represent a certain mechanism of growth but it is still a *single* mechanism.

We have created an unusual and perhaps puzzling distribution but it would be incorrect to be so mesmerised by this simple mathematical operation as to propose different regimes of growth for different perceived parts of the $f(t)/g(t)$ ratio. We can see that the features observed for the GDP/cap distributions can be easily replicated by dividing two mathematically-defined hyperbolic distributions. It is, therefore, clear that hasty assumptions about different socio-economic conditions for the different perceived "parts" of the GDP/cap distributions can be questioned, which means that the whole Unified Growth Theory based on such assumptions can be also questioned.

The next step in explaining the GDP/cap distributions is now to explain why the division of two hyperbolic distributions generates such a puzzling trajectory, which appears to be made of two distinctly different components and why these apparently different components are so strongly pronounced.

**Explaining the ratio of hyperbolic distributions**

Using the eqns (1) and (2) we can see that the ratio of two hyperbolic distributions can be represented also in two other ways:

$$\frac{f(t)[Hyperbolic]}{g(t)[Hyperbolic]} = [g(t)]^{-1}[Linear] \cdot f(t)[Hyperbolic] = \frac{[g(t)]^{-1}[Linear]}{[f(t)]^{-1}[Linear]} \qquad (3).$$

These operations are represented graphically in Figure 3. We can see that all these mathematical operations create the same distribution representing the ratio $f(t)/g(t)$. It does not matter which pathway we take – results are the same.

Dividing two monotonically-increasing hyperbolic distributions is the same as multiplying hyperbolic distribution by a decreasing linear function and the same as dividing two decreasing linear functions. It is all just as simple as that. There are no hidden mysteries that need to be explained by some kind of complicated theories and mechanisms, but we still want to understand why these simple operations generate such a peculiar distribution, which appears to be made of two distinctly different components: horizontal and vertical.

The easiest way to understand the division of hyperbolic distributions is probably by looking at the middle section of Figure 3. The effect of the multiplication of hyperbolic distribution by the decreasing linear function is to lift up the left-hand part of the slowly increasing section of hyperbolic distribution and suppress the right-hand part. However, if $f(t)$ escapes to infinity earlier than $g(t)$, $f(t)$ will be escaping to infinity when $[g(t)]^{-1}$ is still positive. The values of $[g(t)]^{-1}$ will be small but the multiplication of the rapidly increasing values of $f(t)$ by small values of $[g(t)]^{-1}$ will have no effect on the escape to infinity. The product of such numbers will be also rapidly escaping to infinity. The combined effect of such multiplication of a decreasing straight line by the increasing hyperbolic distribution is to flatten the slowly increasing section of the hyperbolic distribution without significantly changing the large values. The initial slow increase is made even slower and the *perceived* transition to the steep part is even more pronounced. However, *there is no mathematically-defined transition at any time*.

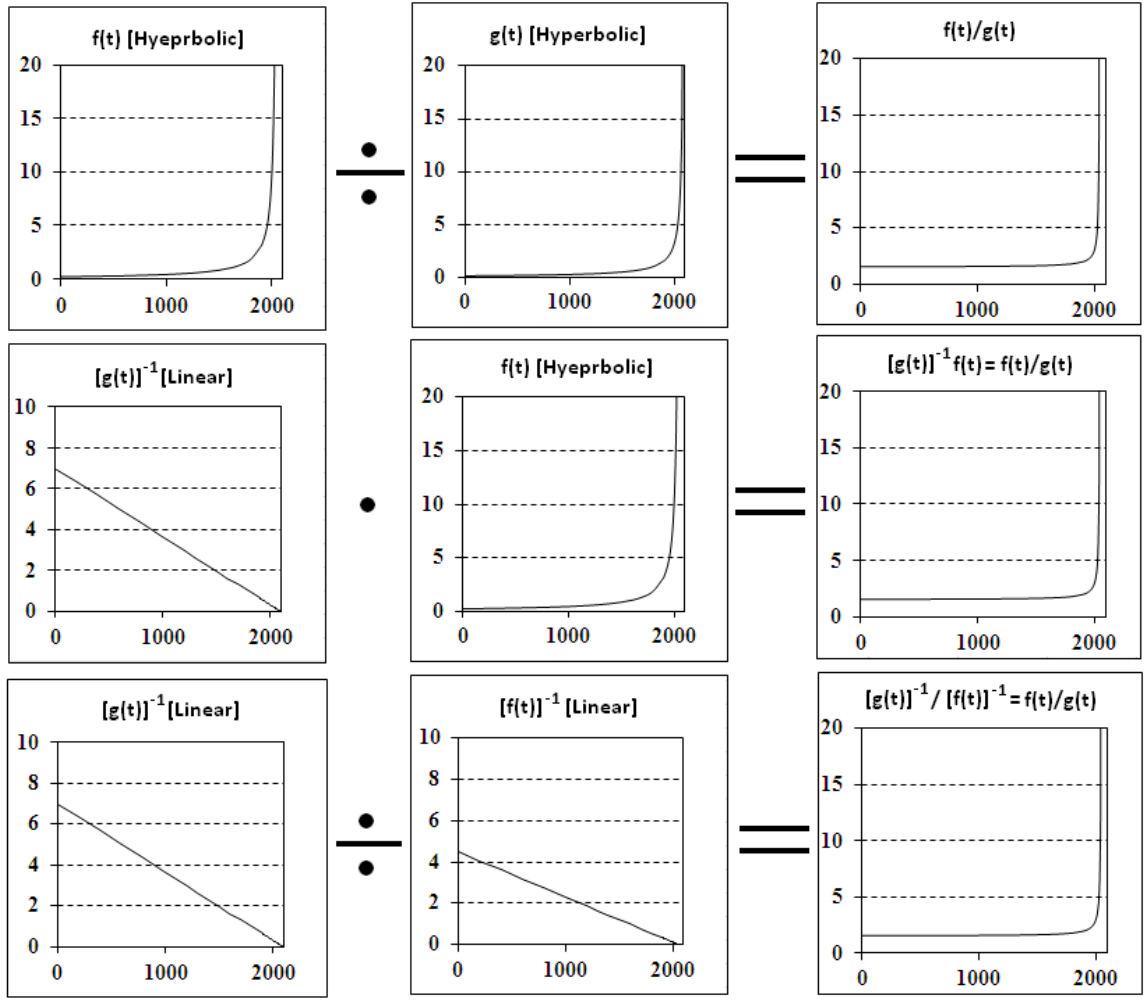

Figure 3. Graphic representation of the eqns (3).

The ratio of two hyperbolic distributions can be described simply as *the linearly-modulated hyperbolic distribution*. Thus in our example the ratio of $f(t)/g(t)$ can be described as *the linearly-modulated hyperbolic $f(t)$ distribution*. The linear modulation is done by the linear function $[g(t)]^{-1}$ representing the reciprocal values of the hyperbolic $g(t)$ distribution.

Likewise, the distribution representing the historical GDP/cap growth can be described as *the linearly-modulated hyperbolic GDP distribution*. The linear modulation is done by the linear distribution representing the reciprocal values of the hyperbolic distribution describing the growth of human population.

The ratio of two hyperbolic distributions looks as if being made of two different components, slow and fast, but it is still the same, uninterrupted, monotonically increasing distribution. It is still a *single* mathematical distribution. It is the distribution, which is not made of two different sections. It is the distribution that it is *impossible* to divide into two distinctly different parts represented by two different functions. This distribution increases slowly over a long time and fast over a short time but the transition from the perceived slow to the perceived fast growth occurs *over the entire range of time*. It is *impossible* to determine the time of this perceived transition. It is impossible to determine the time of the perceived takeoff because *the takeoff does not exist* even if it appears to exist. The perceived takeoff is an illusion. There *is* a slow growth over a long time and a fast growth over a short time but there is no transition at any time between the slow and the fast growth. The slow and the fast growth are represented by the same, monotonically increasing distribution, which is not made of distinctly different components.

Even though the ratio of hyperbolic distributions, $f(t)/g(t)$, looks as if it were made of two or three components (see Figures 2 and 3), even though the distribution represented by this ratio increases slowly over a long time and fast over a short time, even though it increases to infinity at a fixed time and even though it appears to be characterised by a takeoff at a certain time, it is still just a single, monotonically-increasing distribution, which is *impossible* to divide into different components. We have to accept it and learn to live with it.

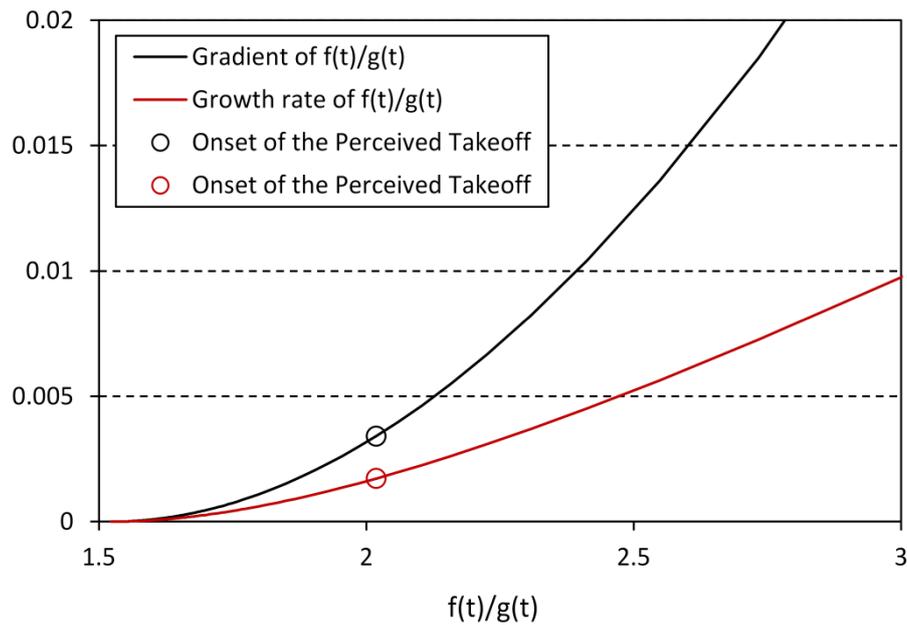

Figure 4. The gradient and growth rate of the ratio of hyperbolic distributions $f(t)/g(t)$. The onset of the perceived takeoff shown in Figure 2 is indicated. This figure shows that the takeoff never happened and that the distribution representing the ratio $f(t)/g(t)$ is not made of different components. It is a single, monotonically-increasing distribution.

The easiest way to understand this apparent paradox and to dispel the strong illusion of different components is to examine the reciprocal values of each of the two components of the ratio $f(t)/g(t)$, i.e. the reciprocal values $[f(t)]^{-1}$ and $[g(t)]^{-1}$ as shown in the lowest part of Figure 3. It would be obviously unreasonable to claim that each of these straight lines is made of two or three distinctly different components and it would be unreasonable to claim different mechanisms of growth for various, arbitrarily-selected parts of a straight line. At which point located on the straight line one mechanism of growth is supposed to end and a new mechanism to begin? It is *impossible* to claim two or three distinctly different sections on the monotonically decreasing straight lines. There is also obviously no feature on such straight lines that could be claimed as marking a takeoff.

We can also take a different approach and demonstrate that the ratio $f(t)/g(t)$ represents a single, monotonically-increasing distribution and that there is no takeoff at any time. This different approach consists in calculating the gradient and the growth rate of the $f(t)/g(t)$ ratio. Results are presented in Figure 4 around the time of the perceived takeoff, i.e. when the $f(t)/g(t)$ reaches the value of 2 (see Figure 2). For better clarity, results are plotted as a function of the size of the $f(t)/g(t)$ ratio.

These calculations show clearly that both the gradient and the growth rate of the hyperbolic ratio $f(t)/g(t)$ increase monotonically. The perceived takeoff never happened. What looks like a takeoff in Figure 2 is in fact just the continuation of the undisturbed and monotonically-increasing distribution representing the $f(t)/g(t)$ ratio. It is impossible to claim different components for any of the distributions displayed in Figure 4, representing the $f(t)/g(t)$ distribution, which in Figure 2 looks very deceptively as being made of two different components. It is impossible to claim a takeoff for any these two distributions. The two components simply do not exist and the takeoff is just an illusion.

**Analysis of the historical GDP/cap data**

The GDP and population data (Maddison, 2001) [the same data as used but not analysed during the formulation of the Unified Growth Theory (Galor, 2005a, 2011)] together with their fitted hyperbolic distributions are shown in Figure 5. Indicated in the figure is the time of the Industrial Revolution 1760-1840 (Floud & McCloskey, 1994), which is generally claimed as the time of the alleged takeoff in the economic growth (Galor, 2005a, 2008a, 2011, 2012). Parameters fitting the GDP data are: $a = 1.716 \times 10^{-2}$ and $k = 8.671 \times 10^{-6}$ while parameters fitting the population data are $a = 8.724$ and $k = 4.267 \times 10^{-3}$.

Points of singularity are: $t_s \approx 1979$ for the world GDP and $t_s \approx 2045$ for the population data. The point of singularity for the world GDP is before the point of singularity for the growth of the world population. Consequently, the GDP/cap ratio should display the same features as shown in Figure 2 for the $f(t)/g(t)$ ratio and indeed it does. The calculated curve and the data shown in Figure 6 follow a similar distribution as shown in Figure 2.

The point of singularity of the GDP/cap trajectory is determined by the point of singularity for the GDP distribution. If the point of singularity for the GDP trajectory were higher than the point of singularity for the population trajectory, the growth of the GDP/cap would have remained nearly constant over a long time and then it would decrease to zero close to the point of singularity of the population trajectory. The danger of the escape to infinity of the GDP/cap growth would have been avoided. Under suitably-chosen conditions the GDP/cap growth can be made to increase or decrease slowly over a much longer time. If only we understood the mathematics of the GDP/cap distributions early enough we could have perhaps been able to control the economic growth in such a way as to ovoid its currently-experienced rapid and dangerous increase.

The fundamental postulate of the Unified Growth Theory (Galor, 2005a, 2011) is the existence of three, distinctly different regimes of growth governed by distinctly different mechanisms: the Malthusian Regime (or Epoch), the Post-Malthusian Regime and the Sustained-Growth Regime. Galor claims precise timing of these three alleged regimes. The data (Maddison, 2001) he uses extend only down to AD 1, but he claims that the Malthusian Regime commenced in 100,000 BC (Galor 2008a, 2012a). He then claims that this regime ended in AD 1750 for developed countries and in 1900 for less-developed countries. The Post-Malthusian Regime was between 1750 and 1870 for developed countries and from 1900 for less-developed countries. The Sustained-Growth Regime commenced in 1870 and continues until the present time. All these interpretations are not supported by a rigorous and scientific analysis of data. They were prompted by the incorrect interpretations of hyperbolic distributions. It appears also that they were fuelled by a good dose of creative imagination, which was allowed to be uncontrolled by the scientific process of investigation.

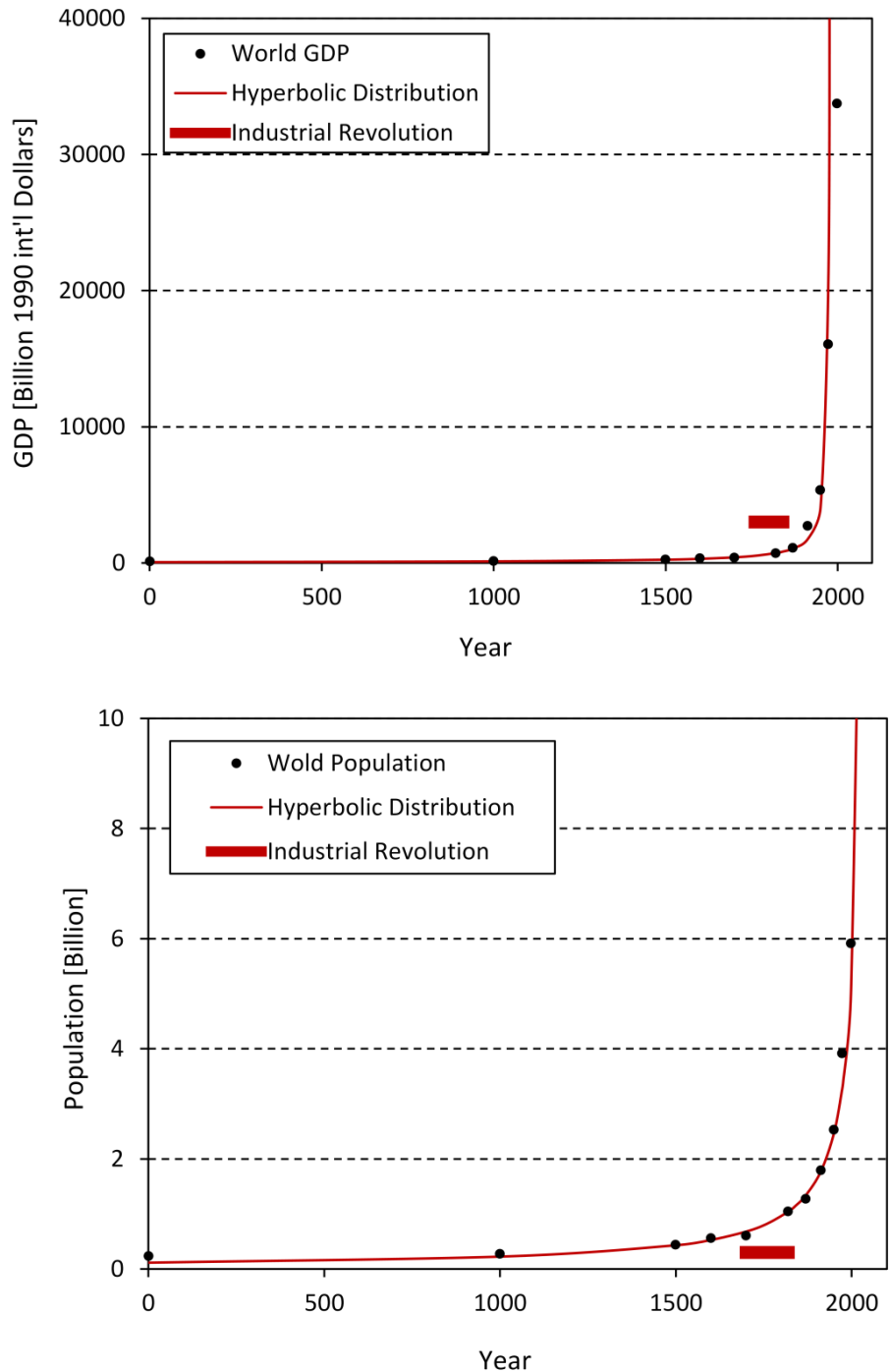

Figure 5. Hyperbolic distributions are compared with the world GDP and population data (Maddison, 2001). The GDP is expressed in billions of 1990 International Geary-Khamis dollars and the population in billions.

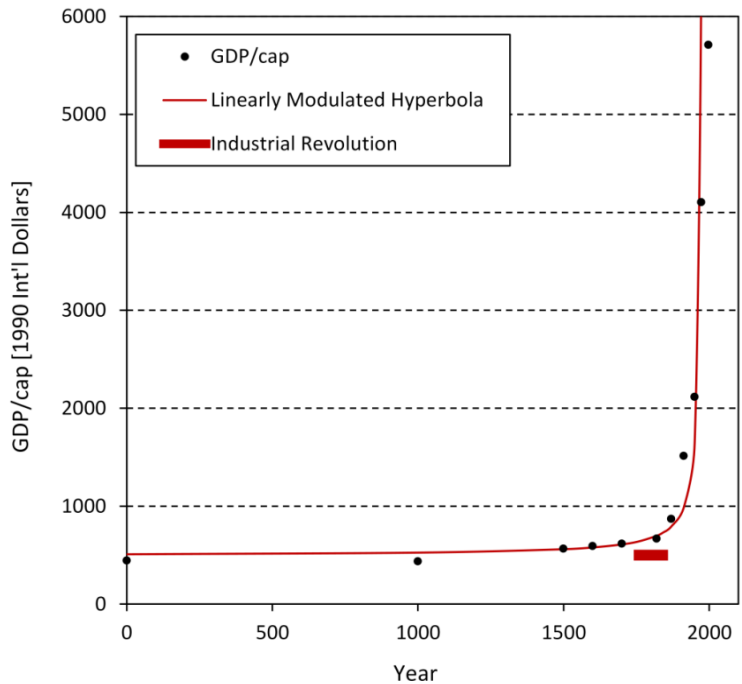

Figure 6. Calculated, linearly-modulated hyperbolic GDP distribution, representing the GDP/cap ratio, is compared with the world GDP/cap data (Maddison, 2001). The GDP/cap is expressed in the 1990 International Geary-Khamis dollars.

If we allow ourselves to be guided by impressions then looking at the data presented in Figure 1 or even in Figure 6 we would have to conclude that there was a long epoch of stagnation followed by a sudden explosion, which Galor describes as a takeoff. This conclusion appears to be obvious and it seems to be supported by the data and by calculated distribution.

However, the fit to the data is not produced by using different mathematical functions describing different perceived stages of growth. If we *had* to use such different functions we could perhaps claim the existence of different regimes of growth but the fit to the data is described by a *single* mathematical distribution.

Furthermore, we already know that there was no takeoff in the growth of the GDP and that the historical GDP trajectory cannot be divided into three different regimes (Nielsen, 2014). This conclusion is based on the examination of the reciprocal values of the GDP data.

The same conclusion applies also to the population data, because, as shown in Figure 5, they are also described by hyperbolic distribution and according to the eqn (2) hyperbolic distributions describing growth are represented by decreasing straight lines. Dividing a straight line into different arbitrary sections clearly makes no sense. Likewise, looking for a change of direction on a straight line to claim a takeoff also makes no sense. Consequently, we can hardly expect that we can produce a takeoff and three different regimes of growth for the GDP/cap ratio.

However, we can prove that there was no takeoff for the GDP/cap distribution and that the three regimes of growth did not exist. We shall do this by calculating the gradient and the growth rate for the calculated GDP/cap trajectory. These calculations are presented in Figures 7 and 8.

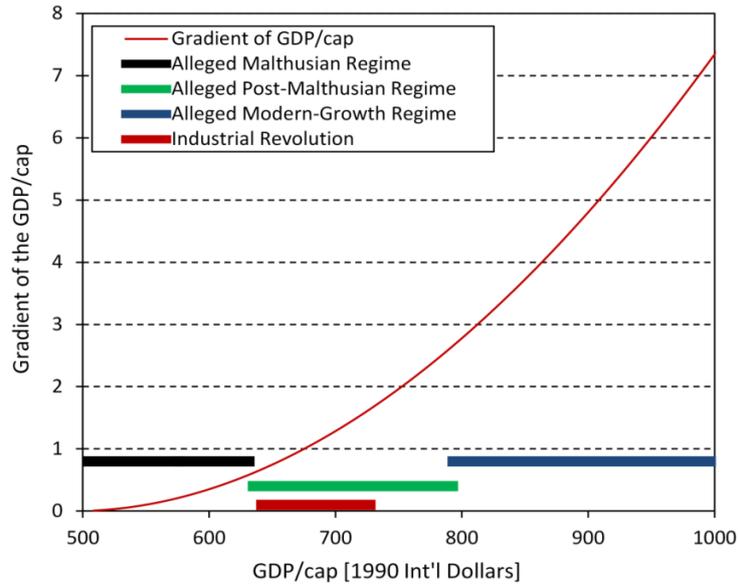

Figure 7. Gradient of the world GDP/cap calculated using the fitted, linearly-modulated hyperbolic distribution shown in Figure 6. The GDP/cap is expressed in the 1990 International Geary-Khamis dollars. There was no takeoff at any time and the three regimes of growth postulated by Galor (2005a, 2011) did not exist.

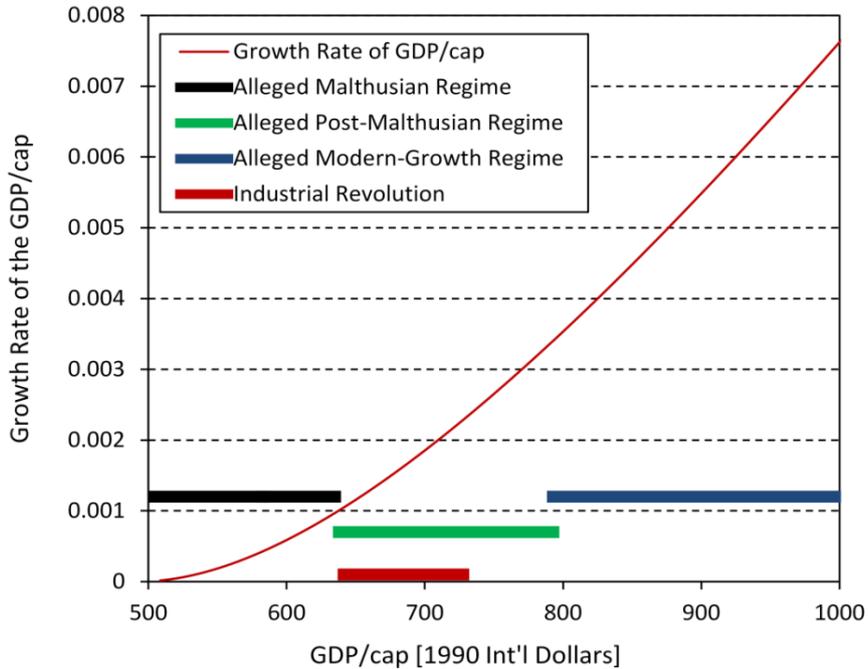

Figure 8. Growth rate of the GDP/cap calculated using the fitted, linearly-modulated hyperbolic distribution shown in Figure 6. The GDP/cap is expressed in the 1990 International Geary-Khamis dollars. There was no takeoff at any time and the three regimes of growth postulated by Galor (2005a, 2011) did not exist.

A takeoff in the GDP/cap trajectory would be marked by a clear change in the gradient and in the growth rate around the time of the Industrial Revolution when a transition to a new economic growth regime was supposed to have happened (Galor, 2005a, 2008a, 2011, 2012a). The shape of the trajectories describing the gradient and growth rate would have to be distinctly different before and after the Industrial Revolution. There should be a certain clear discontinuity.

The gradient and the growth rate of the fitted curve increase monotonically confirming that the fitted, linearly-modulated hyperbolic distribution increases also monotonically. The calculated curve gives excellent fit to the GDP/cap data and consequently the gradient and the growth rate of the fitted curve represent also the gradient and the growth rate of the data.

Figures 7 and 8 clearly demonstrate that there is no reason for terminating the alleged Malthusian Regime around AD 1750 and starting a new regime because there was no unusual change in the gradient and in the growth rate of the GDP/cap around that time, but there was also no scientifically-justified reason for assuming the existence of the Malthusian Regime. There is no reason for terminating the alleged Post-Malthusian Regime around 1870 and starting the alleged Sustained-Growth Regime. There is no reason for chopping the monotonically-increasing distributions into three arbitrarily-selected sections. There is no reason for proposing three regimes of growth governed by distinctly different mechanism. There is no reason for claiming a takeoff at any time.

These calculations, supported by data, clearly demonstrate that the Industrial Revolution had no impact on the economic growth trajectory. Impacts were of different kind but the data show that the Industrial Revolution did not boost the global economic growth. It did not even boost the economic growth in Western Europe (Nielsen, 2014). The three regimes of growth did not exist. The fundamental postulates of the Unified Growth Theory (Galor, 2005a, 2011) are contradicted by the analysis of data, the same data as used but not analysed during the formulation of this theory. Unified Growth Theory describes and explains phenomena that did not exist and consequently it does not explain the historical economic growth. It is an incorrect and misleading theory.

The discussion of socio-economic conditions presented by Galor might be interesting for another reason but there is no evidence in the GDP data that this discussion has any relevance for explaining the mechanism of the economic growth. However, even this discussion, translated repeatedly into mathematical expressions, appears to be dubious and could be also questioned.

Economic growth was indeed slow over a long time and fast over a short time but it is incorrect to divide this monotonically increasing distribution into three regimes and claim distinctly different mechanisms for the arbitrarily selected sections. It is also incorrect to claim that there was a takeoff at a certain time. The data and their analysis give no scientific basis for such claims.

Historical economic growth has to be explained using a *single mechanism*. Such a mechanism should describe the slow and fast growth including the apparent transition. All these "parts" should be treated as one. Only then we could claim that we have explained the mechanism of the historical economic growth.

Dividing the past growth into three different regimes and claiming three different mechanisms is unsupported by data and it does not explain the mechanism of the historical economic growth. A truly unified growth theory will have to be based on a *single mechanism*. Such an explanation will be proposed in a separate publication (Nielsen, 2015c).

**Summary and conclusions**

The aim of our discussion was to explain the puzzling features of the GDP/cap distributions showing a slow growth over a long time, followed by a rapid increase, the features which appear to be causing a significant problem with their interpretations, and the outstanding example is the Unified Growth Theory (2005a, 2011). Our discussion was based on precisely the same data which were used, but not analysed, during the formulation of the Unified Growth Theory. These data represent the *historical* economic growth and the *historical* growth of human population. *They do not represent the current growth* but it is the historical growth that is causing such a big problem.

Historical economic growth, global and regional, shows a clear preference for increasing along hyperbolic trajectories (Nielsen, 2014, 2015a, 2015b). Hyperbolic growth contains singularity, when a growing entity escapes to infinity at a fixed time. However, such a growth becomes, at a certain stage, impossible and has to be terminated either by catastrophic collapse or by a diversion to a slower trajectory, which is hardly surprising because many other types of growth can be and are terminated. For instance, the best known exponential growth, which does not increase to infinity at a fixed time, becomes impossible after a certain time and has to be terminated. Historical economic growth was diverted to slower trajectories between around the end of the 1800s and the mid-1900s (Nielsen, 2015a, 2015b).

Data (Maddison, 2001, 2010) show that the historical hyperbolic growth continued for hundreds of years (Nielsen, 2015a, 2015b) and we have to accept this evidence because the hyperbolic growth is uniquely identified by its reciprocal values. It is remarkable that such an "impossible" type of growth continued for such a long time. However, the mechanism of growth can change and there is nothing unusual about it. The mechanism of the economic growth must have changed because while it is no longer following hyperbolic trend it continues to increase along slower trajectories (Nielsen, 2015a, 2015b). The singularity has been bypassed and the current economic growth and the growth of human population are controlled by different mechanism than in the past.

We have discussed mathematical properties of the historical GDP/cap distributions. We have explained how they should be analysed and interpreted.

If both components of the GDP/cap indicator increase hyperbolically (as for the historical world economic growth and for the growth of the population) then the GDP/cap distributions represent a ratio of hyperbolic trajectories. Created features may be easily confusing and consequently historical GDP/cap data have to be analysed with care. Interpretations based on impressions and supported by the frequently-used crude display of data (Ashref, 2009; Galor, 2005a, 2005b, 2007, 2008a, 2008b, 2008c, 2010, 2011, 2012a, 2012b, 2012c; Galor and Moav, 2002; Snowdon & Galor, 2008) represent a perfect prescription for drawing incorrect conclusions.

We have explained how to understand the confusing features of the historical GDP/cap distributions. They can be interpreted simply as the linearly-modulated hyperbolic GDP distributions. Linear modulation is by the reciprocal values of population data. We have discussed a few ways these distributions can be analysed to understand their characteristic features.

As an illustration of our discussion, we have investigated the data (Maddison, 2001) used in developing the Unified Growth Theory (Galor, 2005a, 2011). Earlier investigation (Nielsen, 2014) of the GDP data (Maddison, 2001) revealed that the fundamental postulates of this theory are unsupported. Now, this conclusion has been confirmed and reinforced by the analysis of the GDP/cap data.

In his theory, Galor discusses various socio-economic concepts of growth but his theory does not explain the mechanism of economic growth because it is based firmly on the misinterpretation of the purely mathematical features of hyperbolic distributions. His discussion of socio-economic issues is interesting but it has no relevance for explaining the mechanism of the economic growth because changes in socio-economic conditions had no effect on the economic growth trajectory as manifested by the available data (Maddison, 2001), the same data, which were used, but not analysed, in the formulation of the Unified Growth Theory. Galor's speculations about socio-economic processes are strongly guided by phantom features created by hyperbolic illusions.

We have demonstrated that the features interpreted in this theory as different stages of growth represent in fact just the *uniform mathematical properties* of the GDP/cap distributions describing a single-stage economic growth. *Unified Growth Theory does not explain the historical economic growth* because the three regimes of growth claimed by this theory did not exist and there was no takeoff at any time. The theory describes features, which do not characterise economic growth.

The data (Maddison, 2001) show that the economic growth was indeed slow over a long time and fast over a short time but the close analysis of these data reveals a *single*, monotonically-increasing distribution. There is no need and no justification for breaking the historical economic growth into different stages of growth. Mathematical analysis of historical economic growth demonstrates that they have to be explained using a *single* mechanism.